\documentstyle[pra,aps,epsfig]{revtex}
\begin{document}
\draft \wideabs{
\title{Storing unitary operators in quantum states}
\author{Jaehyun~Kim, Yongwook~Cheong, Jae-Seung~Lee, and Soonchil~Lee}
\address{Department of Physics, Korea Advanced Institute of Science %
and Technology, Taejon 305-701, Korea}
\date{\today}
\maketitle

\begin{abstract}
We present a scheme to store unitary operators with self-inverse generators in
quantum states and a general circuit to retrieve them with definite success
probability. The continuous variable of the operator is stored in a
single-qubit state and the information about the kind of the operator is
stored in classical states with finite dimension. The probability of
successful retrieval is always 1/2 irrespective of the kind of the operator,
which is proved to be maximum. In case of failure, the result can be corrected
with additional quantum states. The retrieving circuit is almost as simple as
that which handles only the single-qubit rotations and \textsc{cnot} as the
basic operations. An interactive way to transfer quantum dynamics, that is, to
distribute naturally copy-protected programs for quantum computers is also
presented using this scheme.
\end{abstract}

 }
Quantum computers store information in quantum states and process it quantum
mechanically, which make it possible to solve certain problems much faster than
classical computers. The quantum states containing information are generally
in superposed or entangled states that have no classical analog, and the
quantum-mechanical processing is realized by unitary operations while the
processing of a (reversible) classical computer is limited to permutation
operations. Recently, schemes to store the processing operation itself in
quantum states have been proposed by Preskill~\cite{preskill} in the context
of the fault-tolerant quantum computation and by Vidal {\it et
al.}~\cite{vidal} from the point of view of the programmable gate array. In
other words, quantum dynamics is stored in quantum states. Storing and
retrieving an operation means that a unitary operator $U$, which will be
applied to a state $|d\rangle$, is encoded into a state $|U\rangle$, and some
operator $G$ makes the whole state evolve as $G(|U\rangle\otimes
|d\rangle)=|U'\rangle\otimes (U|d\rangle)$, where $|U'\rangle$ is some residual
state. The most distinguished difference between the program stored in quantum
states and the one stored in classical states is that the former is naturally
protected from copy or even reading.

The previous works proposed a way to store only the single-qubit rotation
about $z$ axis in a quantum state and retrieve the operation with definite
success probability. It might be interesting to see what other operations can
be stored in quantum states and how they can be handled. Implementation of
operations would be easier when one can store and transfer several different
kinds of operations than when one can handle only a few basic operations and
the operation of interest has to be decomposed first into a sequence of those
basic operations. In this work, we generalize the scheme to store and retrieve
arbitrary unitary operators satisfying some conditions and present a way to
transfer them.

An arbitrary unitary operator $U_B(\theta)$ can be written as
\begin{equation}
U_B(\theta) = \exp[-\imath (\theta/2) B],  \label{eq-bbb}
\end{equation}
where the generator $B$ is a Hermitian operator of arbitrary dimension, and
$\theta$ is a real number. The number of distinct unitary operators is
infinite because $\theta$ is arbitrary and there are infinitely many different
generators. Therefore, it seems that infinitely many resources are required to
store arbitrary operators~\cite{nielsen}. The number of different kinds of
generators we need to handle is finite because the combinations of basic
operators, such as the single-qubit rotations and controlled-\textsc{not}
(\textsc{cnot}), can make arbitrary unitary operators as well
known~\cite{barenco}. Therefore, the whole point of storing a unitary operator
is to store a continuous variable $\theta$ in quantum states. Storing a real
number in a digitized state requires an infinitely large resource, whether it
is a quantum or classical system. Since quantum system has both the digital
and analog characteristics, however, it is possible to store unitary operators
in finite resources as shown below, only if we allow the possibility of
failure when retrieving. A finite analog system is, in principle, capable of
storing real numbers, though there is the question of precision in the
operation and measurement in practice.

Consider an operator $U_B(\theta)$ of an {\it a priori} known $B$. If $B$ is not
only Hermitian but also unitary, $B$ is self-inverse or $B^2=E$ where $E$ is a
unity operator. Product operators~\cite{sorensen} including Pauli operators
belong to this case, and any Hermitian operator can be expressed as a linear
combination of them. Then, $U_B(\theta)$ is decomposed as
\begin{equation}
U_B(\theta) = \cos(\theta/2)E-\imath\sin(\theta/2)B.
\end{equation}
Here, $U_B(\theta)$ is expressed as a linear combination of two different
operators, $E$ and $B$, and the information about $\theta$ is included in the
coefficients. This expression suggests that $U_B(\theta)$ can be stored in a
single-qubit quantum state, say an {\it angle state}, defined by
\begin{equation}
|\theta\rangle \equiv \cos(\theta/2)|0\rangle -\imath\sin(\theta/2)|1\rangle .
\label{eq-p}
\end{equation}
That is, two operators $E$ and $B$ are mapped onto the two states
$|0\rangle$ and $|1\rangle$, respectively, and the coefficients containing
$\theta$ remain same. Note that this angle state contains only the information
about $\theta$, not about $B$ at all. The information about the
mapping of the operators can be stored in additional
qubits, as will be discussed soon.

\begin{figure}
\epsfig{file=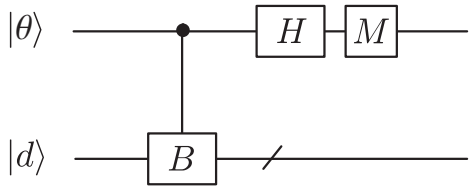,width=8.5cm} \caption{The general gate array $G_B$ that
retrieves $U_B(\theta)$ of a fixed generator $B$ from the angle state
$|\theta\rangle$. $H$ is a Hadamard operator and $M$ represents a projective
measurement in the basis $\{|0\rangle, |1\rangle\}$. The short diagonal line
on the bottom line represents that the data state $|d\rangle$ consists of
several qubits.} \label{fig-1}
\end{figure}

The operator $U_B(\theta)$ is retrieved from the angle state by using a gate
array $G_B$ that consists of a controlled-$B$ defined by
\begin{equation}
|0\rangle\langle 0|\otimes E +|1\rangle\langle 1|\otimes B, \label{eq-cb}
\end{equation}
and a single-qubit Hadamard operator $H$ on the angle state as shown in
Fig.~\ref{fig-1}. Since $B$ is assumed to be unitary, the controlled-$B$ is
also a unitary operator and, therefore, implementable. The total dynamics of
the joint state $|\theta\rangle |d\rangle\equiv|\theta\rangle\otimes|d\rangle$,
where the {\it data state} $|d\rangle$ is a multi-qubit state as $E$ and $B$
have arbitrary dimension in general, are described in terms of $G_B$,
\begin{eqnarray}
G_B|\theta\rangle |d\rangle &=& H[\cos(\theta/2)|0\rangle |d\rangle
-\imath\sin(\theta/2)|1\rangle B|d\rangle ]  \nonumber \\
&=& \frac{1}{\sqrt2}[|0\rangle U_B(\theta)|d\rangle +|1\rangle
U_B(-\theta)|d\rangle ] .
\end{eqnarray}
A projective measurement of the angle state in the basis
$\{|0\rangle,|1\rangle\}$ will make the data state collapse into either a
desired state $U_B(\theta)|d\rangle$ or a wrong state $U_B(-\theta)|d\rangle$
with the equal probability. Therefore, $U_B(\theta)$ stored in the angle state
is retrieved in a probabilistic way. In case of failure, the correct state
may be obtained by executing $G_B$ once more with an additional angle state as
discussed in Refs.~\cite{preskill,vidal}. If we measure the angle state after
the execution of $G_B$ on the joint state of a new angle state
$|2\theta\rangle$ and the wrong data state $U_B(-\theta)|d\rangle$, it will
give the desired state or a new wrong state $U_B(-3\theta)|d\rangle$ with the
equal probability, again. This process can be repeated with the angle state
$|2^m\theta\rangle$ ($m=2,3,\cdots$) until we get the right result.
The average number of the
angle states needed for success is given by $\sum_{m=1}^\infty m(1/2)^m = 2$.

The simplest kind of unitary operators is the single-qubit rotation about
$\alpha$ axis which is written as Eq.~(\ref{eq-bbb}) with $B=\sigma_\alpha$
where $\sigma_\alpha$ is a Pauli operator. In the previous schemes, the angle
of rotation was encoded in phase rather than amplitude like ours but they are
essentially equivalent for single-qubit operations. One advantage of storing
the angle information in amplitude is that the simple mapping in Eq. (3) and
the retrieving circuit in Fig. 1 is applicable to any operators having
self-inverse generators. In Ref.~\cite{vidal}, it is proven that the maximum
probability of successfully retrieving the $z$-rotation stored in a
single-qubit state is 1/2. It is straightforward to extend the proof to the
general unitary operators with self-inverse generators in our scheme (see
Appendix). Therefore, there exists no scheme with higher probability of
retrieval success than ours to store general unitary operators in a
single-qubit state.

\begin{figure}
\epsfig{file=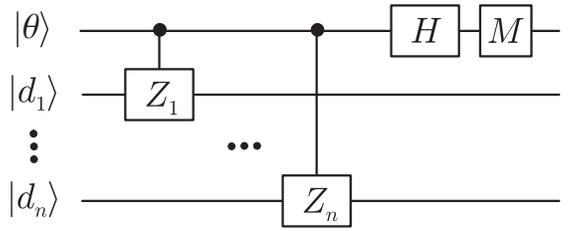,width=8.5cm} \caption{The gate array $G_{12\cdots n}$
for the $n$-bit coupling operator $J_{12\cdots n}(\theta)$. $Z_i$ represents
$\sigma_z$ for the $i$-th data qubit.} \label{fig-2}
\end{figure}

It would be helpful to discuss what useful operators can be stored in the
angle state and present concrete circuits for the gate arrays to retrieve these
operators. Consider an $n$-bit {\it coupling operator} given by
\begin{equation}
J_{12\cdots n}(\theta) =
\exp(-\imath\theta\sigma_{1z}\sigma_{2z}\cdots\sigma_{nz}/2), \label{eq-j}
\end{equation}
which belongs to the kind of operators of our interest because
$(\sigma_{1z}\sigma_{2z}\cdots\sigma_{nz})^2=E$. The corresponding gate array
$G_{12\cdots n}$ is illustrated in Fig.~\ref{fig-2}. This circuit consists of
only $n$ two-qubit operators and a single-qubit Hadamard operator because the
controlled-$\sigma_{1z}\sigma_{2z}\cdots\sigma_{nz}$ is equivalent to the $n$
controlled-$\sigma_{iz}$ ($i=1,2,\cdots,n$). Therefore, as can be seen in
Fig.~\ref{fig-2}, there are direct interactions (vertical solid lines) only
between the angle qubit and each data qubit and the data qubits have no
interactions among them, though $J_{12\cdots n}(\theta)$ itself includes
interaction among all qubits. This nice feature would be useful for the
quantum computer using a special ``head qubit'', which moves to mediate
interactions between non-interacting qubits~\cite{cirac}.

One of the important operators in quantum logic algebra is the {\it controlled
gate} such as the Toffoli gate and phase-shift gate~\cite{nc}. For example, an
$n$-bit phase-shift gate ${\rm diag}[1,1,\cdots,1,e^{\imath\theta}]$ is used
in the quantum factoring algorithm. Consider the operator,
\begin{equation}
\exp\left[-\imath\frac{\theta}{2} \left(P^-_1\! \cdots P^-_{n-1} \sigma_{nz}
+\!\!\!\! \sum_{\alpha_1\!\cdots\alpha_{n-1}}\!\!\!\! P^{\alpha_1}_1\cdots
P^{\alpha_{n-1}}_{n-1}E_n \right) \right],
\end{equation}
where $\alpha_i$ denotes $+$ or $-$, and $P^\pm_i$ are the projection
operators $(E_i\pm\sigma_{iz})/2$ of the $i$-th qubit, respectively. In the
summation, the case of all $\alpha_i$'s being minus is excluded. This operator
is equivalent to $\exp(-\imath\theta\sigma_{nz}/2)$ if all the first $(n-1)$
data qubits are in the state $|1\rangle$ and $e^{-\imath\theta/2}E_n$,
otherwise. Therefore, this is equivalent to the phase-shift gate up to an
overall phase. Since $(P^\pm_i)^2=P^\pm_i$, $P^+_i+P^-_i=E_i$, and $P^+_iP^-_i
=P^-_iP^+_i = 0$, the generator satisfies the self-inverse condition.
Therefore, this operator can be stored in the angle state with the
corresponding controlled-$B$ given by an $n$-bit controlled-$\sigma_{nz}$
which takes the angle qubit and $(n-1)$ data qubits as control bits and
applies $\sigma_z$ to the $n$-th data qubit depending on the state of the
control bits.

We have described how a unitary operator with a self-inverse generator can be
stored in quantum states, and given the corresponding gate array $G_B$ that
retrieves the operator from the quantum states. The angle state contains the
information about $\theta$, and the information about the generator $B$ is
included in the circuit $G_B$ as the controlled-$B$. The information about the
generator is prerequisite for constructing the circuit. Instead of this
impractical design, we can store the information about the generator somewhere
else and construct a general circuit that interprets that information and
execute the corresponding $G_B$. Number of generators to consider is finite
for finite number of data qubits.

One simple way of constructing the general retrieving circuit is to handle the
set of the coupling operators $J_{ij\cdots k}(\theta)$ defined by
\begin{equation}
J_{ij\cdots k}(\theta) =
\exp(-\imath\theta\sigma_{iz}\sigma_{jz}\cdots\sigma_{kz}/2),
\end{equation}
where the generator includes a subset of $n$ spin operators in general. These
operators include $\exp(-\imath\theta\sigma_{iz}/2)$ and $J_{12\cdots
n}(\theta)$ in Eq.~(\ref{eq-j}), and make complete set with \textsc{not} gates
to produce any unitary operators such as the phase shift gate in Eq. (7).
There are $\sum_{m=1}^n \mbox{}_n{\rm C}_m = (2^n-1)$ different generators of
this kind. They can be stored in a {\it command state} consisting of
$\log_22^n=n$ qubits by mapping the generators onto the eigenstates of the
command state. Since the mapping onto the eigenstates is equivalent to using
classical bits, one qubit for the angle state and $n$ classical bits for the
command state are all that required to store all of these coupling operators.

\begin{figure}
\epsfig{file=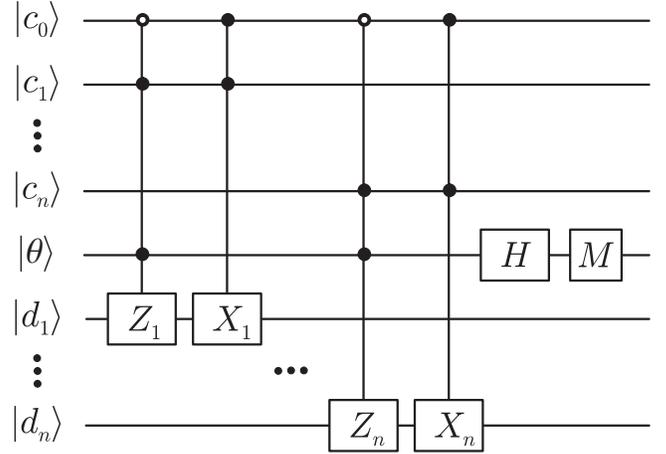,width=8.5cm} \caption{The general gate array handling a
set of basic operators, $J_{ij\cdots k}(\theta)$ and $X_i$.} \label{fig-3}
\end{figure}

The general gate array in Fig.~\ref{fig-3} is slightly modified from the gate
array $G_{12\cdots n}$ in Fig.~\ref{fig-2} to include the command state. For
example, if $J_{1n}(\theta)$ is required to be stored, then the corresponding
controlled-$B$ is the multiplication of $\sigma_{1z}$ and $\sigma_{nz}$ each
of which is controlled by the angle state. Therefore, the angle state contains
$\theta$ and the command state is the binary string $|10\cdots 01\rangle$,
which indicates that only the controlled-$\sigma_{1z}$ and
controlled-$\sigma_{nz}$ are to be activated. To make the circuit complete,
$X=\sigma_{x}$ (\textsc{not} gate) should be added per each qubit. Since $X$
is a fixed operator, one can easily include them in the scheme by employing one
more classical bit $|c_0\rangle$ in the command state. For decoding, $X_i$ is
controlled by $|c_0\rangle$ as well as $|c_i\rangle$ ($i=1,2,\cdots,n$), which
is nothing but the Toffoli gate operation. Consequently, $(n+1)$ classical bits
and one qubit are used to store any operators in the form of $J_{ij\cdots
k}(\theta)$ and $X_i$ which can build up arbitrary unitary operators. The
circuit is almost as simple as that which handles only the single-qubit
rotations and \textsc{cnot}.

This scheme of storing and retrieving quantum operations can be used to
distribute naturally copy-protected programs for quantum computers. A
programming of an quantum algorithm means the process of decomposing the
unitary operation required by the algorithm into a sequence of basic
operations. The more basic operations we have, the easier to program. A
program can be stored either in classical states or quantum states. If a
program is stored in quantum states, it can neither be copied nor read and only
probabilistically retrievable. One way to distribute a quantum program is as
follows. Suppose that Alice has her operator programed in the form of a
sequence of the basic operators. Then, (i) Alice stores the first basic
operator of the sequence in the angle and command states, and sends them to
Bob, (ii) Bob performs the gate array of Fig.~\ref{fig-3}, and tell Alice the
measurement result -- whether the operation has succeeded or not, (iii) Alice
sends to Bob the new angle state in case of failure or the next operator of the
sequence when succeeded, and (iv) they repeat (ii) and (iii) until the last
operator of the sequence is transferred. Although there is possibility of
failure in each operation transfer, it can always be corrected and the average
number of the angle states necessary for successful operation is only two.
This is an interactive distribution scheme where Alice and Bob have to
communicate with each other during the transfer to guarantee the successful
operation.

In conclusion, we have shown that the unitary operators with self-inverse
generators can be stored in a quantum state and retrieved exactly by encoding
the continuous variable into the probability amplitude of the state, at the
cost of possible failure. This probabilistic feature is the cost we have to pay
to store a real number in a quantum state of finite dimension, even which is
impossible in classical systems. Utilizing the circuits for the coupling
operators and storing the information about the generators in classical states
of finite size, it is possible to store and retrieve arbitrary operators
(including such ones having non-self-inverse generators), and the general
retrieving circuit is very simple.

This work was supported by NRL Program, electron spin science center, and
BK21 Project.

\appendix
\section*{}
Any scheme to store the quantum operation $U(\theta)=\exp[+\imath (\theta/2)B]$
of a fixed, self-inverse generator $B$ in the quantum states can be described
by a unitary transformation,
\begin{equation}
G (|U_\theta\rangle\otimes |d\rangle )= \sqrt{p^d_\theta} |\tau^d_\theta\rangle
\otimes U(\theta)|d\rangle  + \sqrt{1-p^d_\theta} |\chi^d_\theta\rangle ,
\label{eq-a1}
\end{equation}
where $|U_\theta\rangle$ is the {\it program state} containing the information
about the operation and $|d\rangle$ is the data state. $G$ transforms the
total state into the sum of the product of some residual state
$|\tau^d_\theta\rangle$ and the desired state $U(\theta) |d\rangle$ with
success probability $p^d_\theta$ and the failed state $|\chi^d_\theta\rangle$
with probability $1-p^d_\theta$. To distinguish success from failure by
measurement, it should be satisfied that
$\langle\tau^d_\theta|\chi^{d'}_{\theta'}\rangle=0$ for all $d$, $d'$,
$\theta$, and $\theta'$.

Suppose that $|d\rangle$ is an $n$-qubit state and $N=2^n$. The data state
$|d\rangle$ is expanded by the computational basis as $ |d\rangle =
\sum^{N-1}_{k=0} c_k |k\rangle$, where $c_k$ is the complex coefficient that
satisfies the normalization condition. Then, Eq.~(\ref{eq-a1}) is rewritten as
\begin{eqnarray}
\lefteqn{G\left( \sum c_k |U_\theta\rangle |k\rangle \right)=} \nonumber \\
&&\sum c_k \left[ \sqrt{p^k_\theta} |\tau^k_\theta\rangle U(\theta) |k\rangle
 + \sqrt{1-p^k_\theta} |\chi^k_\theta\rangle \right], \label{eq-a2}
\end{eqnarray}
where product sign $\otimes$ between the program and data states is omitted for
simplicity. This implies that $p^d_\theta$ and $|\tau^d_\theta\rangle$ do not
depend on $|d\rangle$ because RHS's of Eqs.~(\ref{eq-a1}) and (\ref{eq-a2})
must be same for all $|d\rangle$. Now, $p^d_\theta$ and
$|\tau^d_\theta\rangle$ will be denoted by $p_\theta$ and
$|\tau_\theta\rangle$, respectively.

The one-qubit program state can be also expanded by
$|U_\theta\rangle=\alpha(\theta)|\underline{0}\rangle
+\beta(\theta)|\underline{\pi}\rangle$, where $\alpha(\theta)$ and
$\beta(\theta)$ are complex function satisfying the normalization condition
$\langle U_\theta|U_\theta\rangle=1$, and $|\underline{0}\rangle$ and
$|\underline{\pi}\rangle$ are the program states corresponding to the
operations $E$ and $B$, respectively. For any working scheme $G$, there always
exist the program states for $\theta=0$ and $\theta=\pi$. These states
correspond to $E$ and $B$, respectively, because $U(\theta)=\exp[+\imath
(\theta/2)B] = \cos(\theta/2)E+\imath\sin(\theta/2)B$, $U_\theta = E$ for
$\theta=0$, and $U_\theta = B$ for $\theta=\pi$ (up to an overall phase).
Therefore, we can always expand $|U_\theta\rangle$ by the linear combination of
$|\underline{0}\rangle$ and $|\underline{\pi}\rangle$, which are not
necessarily orthonormal to each other. Again, Eq.~(\ref{eq-a1}) is expressed as
\begin{eqnarray}
\lefteqn{G[ \alpha(\theta)|\underline{0}\rangle|d\rangle
+\beta(\theta)|\underline{\pi}\rangle|d\rangle ] =}
\nonumber \\
&& \alpha(\theta)\left[\sqrt{p_0}|\tau_0\rangle E|d\rangle
+\sqrt{1-p_0}|\chi^d_0\rangle \right] \nonumber \\
&& \mbox{}+ \beta(\theta)\left[\sqrt{p_1}|\tau_1\rangle B|d\rangle
+\sqrt{1-p_1}|\chi^d_1\rangle \right],
\end{eqnarray}
for all $|d\rangle$. Therefore, all $|\tau_\theta\rangle$, $|\tau_0\rangle$,
and $|\tau_1\rangle$ are same with $|\tau\rangle$ not depending on
$|d\rangle$, and
\begin{equation}
\sqrt{p_\theta}U(\theta) = \alpha(\theta)\sqrt{p_0}E+\beta(\theta)\sqrt{p_1}B.
\label{eq-al}
\end{equation}
This means that $\alpha(\theta)=\sqrt{p_\theta/p_0}\cos(\theta/2)$ and
$\beta(\theta)=\sqrt{p_\theta/p_1}\sin(\theta/2)$. The remaining part of the
proof is same with that of Ref.~\cite{vidal}. The maximum probability of
success is 1/2, which is achieved by our scheme.

\end{document}